\documentclass[twocolumn,a4paper,10pt]{article}

\usepackage[utf8]{inputenc}
\usepackage[T1]{fontenc}
\usepackage[russian,english]{babel}
\usepackage{amsmath, amssymb, amsthm}
\usepackage{graphicx}
\usepackage{booktabs}
\usepackage{multirow}
\usepackage{hyperref}
\usepackage{cite}
\usepackage{caption}
\usepackage{subcaption}
\usepackage{float}
\usepackage{geometry}
\usepackage{parskip}
\usepackage{enumitem}
\usepackage{tabularx}
\setlist{nosep, left=0pt}

\usepackage{titlesec}
\titleformat{\section}{\large\bfseries}{}{0em}{}
\titleformat{\subsection}{\normalsize\bfseries}{}{0em}{}
\titleformat{\subsubsection}{\small\bfseries}{}{0em}{}

\hypersetup{
    colorlinks=true,
    linkcolor=blue,
    citecolor=blue,
    urlcolor=blue
}
\usepackage{listings}
\usepackage{xcolor}

\usepackage{algorithm}
\usepackage{algpseudocode}

\begin{document}

\title{\textbf{When Does Learning Beat Heuristics? A Case Study in Kubernetes Scheduler Score Plugins}}
\author{
	Wang Xuying \\
	I. Razzakov Kyrgyz State Technical University	 
	\and
	Zhibek Sarypbekova\\
	I. Razzakov Kyrgyz State Technical University 
}
\date{}
\maketitle

\begin{abstract}
\noindent Kubernetes scheduler plugins that score candidate nodes are, in production, hand-tuned heuristics (\texttt{NodeResourcesFit}, \texttt{NodeResourcesBalancedAllocation})~\cite{kubernetes-kep}. We ask whether a learned scoring function --- trained on real placement decisions from a production cluster trace --- can match or exceed these heuristics, and if not, why. We implement \texttt{AIScore}, an external HTTP-backed Score plugin for the \texttt{kube-scheduler-simulator}, and evaluate two learned models (a Random Forest over engineered features, and a GraphSAGE-based encoder~\cite{graphsage} over per-job task-dependency graphs) trained on the Alibaba Cluster Trace v2018~\cite{alibaba-trace}. Using standard regression fit (R²), both models show modest but monotonically improving quality across four feature-engineering iterations, culminating at R² $\approx 0.042$. However, when we instead evaluate the models on the metric that actually matters for scheduling --- Top-1 ranking accuracy, i.e., whether the model assigns the highest score to the machine the production scheduler (Fuxi) actually chose --- both learned models are \textbf{outperformed by a trivial single-feature heuristic} (rank by free CPU: 74--84\% accuracy vs. 65--66\% for either learned model). We show this gap is best explained by an objective mismatch: both models were trained with pointwise regression (MSE) rather than a ranking-specific objective, echoing a long-standing distinction in the learning-to-rank literature~\cite{burges2005, lambda}. This result closely parallels prior evidence that learned, RL-trained schedulers such as Decima~\cite{decima} and DeepRM~\cite{deeprm} can substantially outperform heuristics when the training objective is aligned with the deployment task, and suggests that objective misalignment --- not architecture --- is the primary obstacle in our setting. We further report a systematic ablation of the resource-occupancy reconstruction required to make offline trace data usable at all (naive features yield R² $\approx 0$), a controlled comparison between the Random Forest and GNN models isolating the effect of feature richness and data volume, and a production-oriented sensitivity analysis of inference latency and serving-container memory constraints. All code, data pipelines, and experiment scripts are released for reproducibility.
\end{abstract}


\section{1.Introduction}

Kubernetes' default scheduler ranks feasible nodes for each pod using a weighted sum of plugin scores, each implementing a hand-designed heuristic~\cite{kubernetes-kep}. \texttt{NodeResourcesBalancedAllocation}, for instance, favors nodes where CPU and memory utilization are similar after placement --- a reasonable, general-purpose rule, but one fixed at the time the plugin was written, blind to workload-specific or cluster-specific patterns that a learned model might exploit.

This raises a natural question, revisited periodically in the systems literature --- from flow-based optimal schedulers like Firmament~\cite{firmament} to learned, DAG-aware schedulers like Decima~\cite{decima} (Section~\ref{sec:related}): can a model trained on historical placement data do better? We investigate this question empirically, using \texttt{kube-scheduler-simulator} (kubernetes-sigs) as a lightweight, reproducible testbed, and the Alibaba Cluster Trace v2018~\cite{alibaba-trace} as a source of real production placement decisions.

Our investigation proceeds in four stages, each motivated by the failure or limitation of the previous one:

\begin{enumerate}
    \item We first validate the experimental pipeline on synthetic data (Section~\ref{sec:synthetic}), confirming a custom \texttt{ScorePlugin} can be wired into the simulator and meaningfully affect placement.
    \item We train a Random Forest on real trace data, discovering that a naive feature construction is \textit{uninformative by construction} (R² $\approx 0$) unless node occupancy is reconstructed at the correct point in time via a sweep-line algorithm (Section~\ref{sec:rf}). Four subsequent feature-engineering refinements each yield measurable, monotonic improvement.
    \item We build a graph-structured alternative (GraphSAGE~\cite{graphsage} over per-job DAGs) to test whether the full dependency structure --- not just aggregate degree statistics --- carries additional signal (Section~\ref{sec:gnn}).
    \item Critically, we discover that the standard regression metric (R²) used in Stages 2--3 \textbf{does not correlate with what actually matters for scheduling}: whether the model ranks the real chosen machine above its competitors. Under a Top-1 ranking evaluation, both learned models lose to a one-line heuristic (Section~\ref{sec:ranking}) --- a result that reframes the entire comparison and, we argue, is itself a transferable methodological lesson for ML-for-systems work.
\end{enumerate}

We close with a production-facing sensitivity analysis (latency, memory) that is largely orthogonal to model quality, but necessary for any deployment discussion (Sections~\ref{sec:latency}--\ref{sec:hetero}).

\textbf{Contributions:}
\begin{itemize}
    \item An open-source, reproducible pipeline for training and evaluating custom Kubernetes scheduler score plugins against a real production trace, including the (non-obvious) occupancy-reconstruction step required to make the trace usable at all.
    \item An empirical demonstration that regression-based training metrics can be systematically misleading for scheduling-relevant ranking quality, with a controlled comparison isolating this effect from feature quality and model capacity.
    \item A controlled Random-Forest-vs-GNN comparison that disentangles architecture from data volume and feature richness --- a confound we argue is under-examined in prior comparisons of this kind.
    \item Production-sensitivity measurements (inference latency overhead, memory-constraint robustness) for an HTTP-served ML scoring plugin, of independent practical interest for anyone deploying similar architectures.
\end{itemize}

\section{2.Related Work}
\label{sec:related}

\subsection{2.1.Cluster Scheduling Architectures}
Kubernetes' scheduler exposes an extensible framework of filter and score plugins (\texttt{NodeResourcesFit}, \texttt{NodeResourcesBalancedAllocation}, \texttt{TaintToleration}, etc.), each a hand-designed heuristic, formalized in the upstream Scheduling Framework design proposal~\cite{kubernetes-kep}; \texttt{AIScore} (Section~\ref{sec:methodology}) is implemented as one such drop-in plugin rather than a scheduler replacement. Two influential non-learned cluster schedulers are relevant context: \textbf{Firmament}~\cite{firmament} reduces placement to a min-cost max-flow optimization over a flow network, achieving sub-second placement at 10,000+-machine scale while matching or exceeding the placement quality of several widely-used centralized and distributed schedulers, and was later integrated with Kubernetes as the Poseidon-Firmament scheduler. Firmament represents the high-quality-optimization end of the design space our single-HTTP-call \texttt{AIScore} plugin sits at the opposite (low-latency, per-node-scoring) end of. More recently, \textbf{Pollux}~\cite{pollux} co-adaptively tunes both per-job training configuration and cluster-wide resource allocation for deep-learning workloads using a ``goodput'' objective combining throughput and statistical efficiency, reporting 37--50\% reductions in average job completion time over prior DL-cluster schedulers --- a further data point that objective design, not just model architecture, drives the gains reported by learned schedulers. Two recent surveys --- Rejiba and Chamanara~\cite{rejiba2022} on custom Kubernetes scheduling broadly, and Senjab et al.~\cite{senjab2023} specifically on Kubernetes scheduling algorithms including AI-focused approaches --- position our narrower, empirical contribution (an explicit R²-vs-ranking-accuracy comparison on one plugin) against the wider landscape of proposed Kubernetes scheduler modifications.

\subsection{2.2.Learned Schedulers}
The most directly related prior work is \textbf{Decima}~\cite{decima}, which trains a graph neural network to embed job dependency DAGs and feeds the resulting representation into a reinforcement-learning policy that jointly decides which job to schedule next and how much parallelism to grant it, reporting at least 21\% improvement in average job completion time over hand-tuned heuristics on a 25-node Spark cluster. Our GNN prototype (Sections~\ref{sec:gnn}) uses the same high-level idea --- a graph encoder over job DAGs --- but in a narrower role (scoring individual node--task pairs for a \texttt{ScorePlugin}, rather than jointly deciding scheduling order and resource allocation via RL) and evaluates it specifically against the ranking metric that governs deployment behavior (Section~\ref{sec:ranking}), a comparison we are not aware of being made explicit in prior learned-scheduler evaluations. Earlier RL-based resource managers such as DeepRM~\cite{deeprm} established that policy-gradient RL can outperform heuristics like Shortest-Job-First and Tetris on synthetic packing workloads, predating Decima's DAG-aware extension. \textbf{RLScheduler}~\cite{rlscheduler} extends this line of work to HPC batch scheduling, using a kernel-based neural network and trajectory filtering to learn scheduling policies directly from trial and error, without hand-designed priority functions, and reports stable performance even on unseen workloads --- reinforcing that RL-trained (rather than pointwise-regression-trained) schedulers are the setting in which learned approaches have most convincingly outperformed heuristics to date. Closer to our GNN prototype's architectural choice, Zhao et al.~\cite{zhao2021} use a graph neural network directly for distributed scheduling decisions, encoding jobs and machines as distinct node types --- a design pattern our per-task GraphSAGE encoder (Section~\ref{sec:methodology}) shares, though applied here to per-candidate scoring rather than joint distributed decision-making. \textbf{Lyra}~\cite{lyra} more recently demonstrates that elastic, GPU-sharing-aware scheduling for deep-learning clusters can substantially improve cluster-wide GPU utilization, further illustrating the breadth of scheduling sub-problems (beyond the single-node-scoring problem we study) to which learned or adaptive techniques have been successfully applied.

\subsection{2.3.Cluster Traces for Scheduling Research}
We use the \textbf{Alibaba Cluster Trace v2018} (\texttt{alibaba/clusterdata})~\cite{alibaba-trace}, one of a series of production traces Alibaba has released for cluster-management research. The \textbf{Google Cluster Trace} family (2011 and 2019 releases)~\cite{google2011, borg} is the other major public trace lineage used in this line of research; the 2019 release extends coverage to eight clusters and enables direct comparison of scheduling behavior across Borg deployments, complementing the single-cluster, batch-scheduler-focused view of the Alibaba trace we use here.

\subsection{2.4.Learning-to-Rank}
Our diagnosis in Section~\ref{sec:ranking} --- that a model trained via pointwise regression can underperform on a ranking task it is not directly optimized for --- echoes a foundational distinction in the learning-to-rank literature. RankNet~\cite{burges2005} first framed relevance ordering as a pairwise classification problem trained via cross-entropy on score differences, rather than pointwise regression to an absolute relevance label; LambdaRank and LambdaMART, surveyed in~\cite{lambda}, further shape gradients directly by the ranking-metric impact of swapping a given pair, rather than by a smooth pointwise loss. Our proposed next step (Section~\ref{sec:conclusion}) --- retraining with a pairwise margin loss --- is a direct application of this pairwise framing to the scheduling-placement setting, which, to our knowledge, is not standard practice in the learned-scheduler literature reviewed above (Decima and DeepRM both use RL rather than a supervised ranking loss).

\subsection{2.5.GNNs for Systems Problems}
Beyond Decima's use of a GNN for job-DAG embedding, graph neural networks such as GraphSAGE~\cite{graphsage} have been applied to a range of adjacent systems problems where the input is naturally graph-structured, motivating the DAG-encoder design used in Sections~\ref{sec:gnn} of this paper.

\subsection{2.6.Positioning}
Relative to this body of work, we see this paper's contribution as threefold: (a) unlike Decima and DeepRM, which are evaluated via simulated or live-cluster job-completion-time improvements under an RL training loop, we evaluate directly against real historical placement \textit{decisions} from a production scheduler (Fuxi) recorded in a public trace, isolating the supervised-learning question of whether a model can reproduce those decisions before any RL-based sequencing or resource-allocation logic is layered on top; (b) we explicitly and quantitatively demonstrate the gap between a standard regression-fit metric (R²) and ranking accuracy on the same models and data --- a distinction implicit in the learning-to-rank literature~\cite{burges2005, lambda} but, to our knowledge, not previously demonstrated as a concrete pitfall in the ML-for-scheduling context; and (c) our sweep-line occupancy-reconstruction methodology (Section~\ref{sec:methodology}), needed to make the trace's static snapshots usable for training at all, may be a reusable recipe for other researchers working with the same or structurally similar traces.

\section{3.Methodology}
\label{sec:methodology}

\subsection{3.1.Environment}
All experiments run on \href{https://github.com/kubernetes-sigs/kube-scheduler-simulator}{kube-scheduler-simulator} (kubernetes-sigs), deployed via Docker Compose with KWOK providing a lightweight kube-apiserver/kubelet emulation. We implement a custom \texttt{ScorePlugin}, \texttt{AIScore}, registered into the simulator's debuggable scheduler via \texttt{debuggablescheduler.WithPlugin}, consistent with the extensible plugin design described in the Kubernetes Scheduling Framework proposal~\cite{kubernetes-kep}. On every \texttt{Score()} invocation the plugin performs a synchronous HTTP POST to an external Python (FastAPI) inference service, carrying node- and pod-level features, and receives back an integer score in [0, 100].

\subsection{3.2.Data Source}
We use the \textbf{Alibaba Cluster Trace v2018} (\texttt{alibaba/clusterdata})~\cite{alibaba-trace}, covering $\sim$4000 machines over an 8-day window, including batch task/instance placement decisions made by Alibaba's production Fuxi scheduler. Four tables are used: \texttt{machine\_meta}, \texttt{machine\_usage}, \texttt{batch\_task}, \texttt{batch\_instance}. Per the trace's documented schema, CPU fields use a ``100 units = 1 core'' convention (converted to milliCPU via $\times 10$); memory fields are normalized to [0, 100] relative to an undisclosed reference and are used in this normalized form throughout, since absolute byte values cannot be recovered.

\subsection{3.3.Label Construction as Learning-to-Rank}
The trace records only the machine actually selected for each instance, not scores for the alternatives Fuxi considered. We frame training as \textbf{learning to rank}, in the spirit of~\cite{burges2005, lambda}: the chosen machine is a positive example (label 100); other resource-feasible machines available at the same historical timestamp are negative examples (label 20), restricted (in later iterations) to the same failure domain as the chosen machine, to better approximate Fuxi's plausible candidate set.

\subsection{3.4.Occupancy Reconstruction}
Free node capacity must reflect actual concurrent occupancy at an instance's start time --- not simply (allocatable $-$ this instance's own request). We verified this is not optional: the naive approximation produced a model indistinguishable from noise (R² $\approx 0$; Section~\ref{sec:rf}). We reconstruct true occupancy via a \textbf{sweep-line algorithm}: occupancy-change events (+request at start, $-$request at end) are built per instance--machine pair, sorted by time, and cumulatively summed per machine; a query at any historical timestamp is served via \texttt{pandas.merge\_asof}.

\subsection{3.5.Models}
\begin{itemize}
    \item \textbf{Random Forest} (scikit-learn, 150 trees, max depth 10) over a flat feature vector: reconstructed free CPU/memory, requested CPU/memory, failure-domain identifiers, and (in later iterations) DAG-derived features (in-degree, out-degree, job width, root/leaf indicators) parsed from the trace's documented task-naming convention.
    \item \textbf{GNN prototype}: a 2-layer GraphSAGE~\cite{graphsage} encoder producing per-task embeddings from a job's dependency graph (nodes = tasks, edges = dependencies parsed from \texttt{task\_name}), combined via a 3-layer MLP with the same machine-level features used by the Random Forest. This follows the general graph-encoder design used for job DAGs by Decima~\cite{decima}, applied here to a per-candidate scoring role rather than an RL policy.
\end{itemize}

\subsection{3.6.Evaluation Metrics}
\begin{itemize}
    \item \textbf{R²} on the held-out regression task (both models trained via MSE against the 100/20 label scheme), reported for comparability with standard ML practice.
    \item \textbf{Top-1 ranking accuracy}: for each group of one positive and up to three negative candidates sharing a \texttt{group\_id}, whether the model assigns the positive example the highest score in its group --- the metric that directly answers ``would the model reproduce Fuxi's decision.''
    \item \textbf{Synthetic load-balancing benchmark}: standard-deviation of per-node CPU/memory utilization across a controlled workload (5 pod-size profiles, 3 heterogeneous nodes), used both as an independent sanity check (Section~\ref{sec:synthetic}) and to characterize homogeneous-vs-heterogeneous workload sensitivity (Section~\ref{sec:hetero}).
    \item \textbf{Operational metrics}: HTTP round-trip inference latency (measured inside the Go plugin via \texttt{time.Since()}), and container behavior (pods scheduled, OOM kills) under artificial network delay and memory constraints.
\end{itemize}

\section{4.Results}

\subsection{4.1.Sanity Check: Synthetic Workload}
\label{sec:synthetic}

Before using real trace data, we validate the plugin pipeline end-to-end: \texttt{AIScore}, trained on a synthetic label penalizing post-placement CPU/memory imbalance, is compared against the scheduler with \texttt{AIScore} disabled (default plugins only) on a controlled workload (5 profiles $\times$ 8 instances = 40 pods, 3 heterogeneous nodes).

\begin{table}[H]
    \centering
    \caption{Synthetic workload: balancing stddev comparison.}
    \begin{tabular}{l c c}
        \toprule
        \textbf{Metric} & \textbf{Baseline} & \textbf{AIScore} \\
        \midrule
        CPU utilization, stddev & \textbf{4.54\%} & 4.87\% \\
        Memory utilization, stddev & \textbf{4.56\%} & 5.13\% \\
        \bottomrule
    \end{tabular}
\end{table}
\noindent This is single run: the small baseline advantage is expected, since the synthetic label imitates rather than improves upon the target heuristic.

This confirms the plugin and benchmark harness function correctly and can detect meaningful (if small) differences in placement quality.

\subsection{4.2.Random Forest on Real Trace Data: Feature Ablation}
\label{sec:rf}

We evaluate four progressively refined training configurations, each addressing a specific limitation identified in the previous one:

\begin{table}[H]
	\centering
	\caption{Random Forest feature ablation results.}
	\small
	\begin{tabularx}{0.95\columnwidth}{c X c c}
		\toprule
		\textbf{Variant} & \textbf{Change} & \textbf{Train R²} & \textbf{Test R²} \\
		\midrule
		A & Uniform-random negative sampling; flat resource features only & 0.0250 & 0.0249 \\
		B & + failure-domain identifiers as features & 0.0289 & 0.0285 \\
		C & Negative sampling restricted to chosen machine's failure domain & 0.0331 & 0.0327 \\
		\textbf{D} & \textbf{+ DAG-derived features} (in/out-degree, job width, root/leaf) & \textbf{0.0420} & \textbf{0.0415} \\
		\bottomrule
	\end{tabularx}
\end{table}
\noindent Random Forest, trained on $\sim$2.9M positive examples. These come from a 3M-row sample of \texttt{batch\_instance} (after cleaning, $\sim$2.94M valid records remained), joined against the full \texttt{batch\_task} table for job/DAG metadata. job width ranked 2nd in feature importance (0.170).

A naive occupancy approximation (allocatable $-$ this instance's own request, ignoring concurrently running instances) was tested prior to Variant A and produced R² $\approx 0$ for both train and test splits --- the positive and negative examples were statistically indistinguishable given that feature construction. This motivates the sweep-line reconstruction described in Section~\ref{sec:methodology} as a necessary, not optional, step.

Feature importances for Variant D place \texttt{job\_width} (the number of tasks in the parent job) second overall (0.170), ahead of \texttt{node\_free\_mem} (0.112) --- a cheap, purely structural feature that required no occupancy reconstruction at all, unexpectedly rivaling the resource-based features that motivated most of the engineering effort.

\subsection{4.3.Graph Neural Network: Full DAG Structure vs. Aggregate Statistics}
\label{sec:gnn}

To test whether the full dependency graph carries signal beyond the aggregate degree statistics of Variant D, we parse \texttt{task\_name} into an explicit edge list per job (rather than only degree counts) and train a GraphSAGE~\cite{graphsage} encoder. From a 500K-row \texttt{batch\_task} sample, 71,124 job graphs were constructed (2--127 tasks per job, median 3).

\begin{table}[H]
    \centering
    \caption{GNN vs. Random Forest regression performance.}
    \begin{tabular}{l c}
        \toprule
        \textbf{Model} & \textbf{R²} \\
        \midrule
        Random Forest, Variant D & \textbf{0.0420} \\
        GNN prototype (5 epochs) & 0.021--0.028 \\
        \bottomrule
    \end{tabular}
\end{table}

\noindent The GNN was trained on 72,734 positive examples --- roughly 40$\times$ fewer than Variant D --- using only 2 node features (\texttt{plan\_cpu}, \texttt{plan\_mem}) versus the Random Forest's 11 engineered features.

Taken at face value, this suggests the GNN underperforms. Section~\ref{sec:ranking} shows this comparison is confounded.

\subsection{4.4.The Metric That Matters: Top-1 Ranking Accuracy}
\label{sec:ranking}

R² measures fit to the arbitrary 100/20 label scheme, not whether the model would reproduce Fuxi's actual decision among the real candidates it faced. We compute \textbf{Top-1 accuracy} directly: for each candidate group, does the model's highest score fall on the positive (chosen) machine? We compare against two baselines --- a random score, and a one-line heuristic ranking candidates purely by reconstructed free CPU.

\begin{table}[H]
    \centering
    \caption{Top-1 ranking accuracy comparison.}
    \begin{tabularx}{0.95\columnwidth}{l X X}
        \toprule
        \textbf{Method} & \textbf{RF dataset} (n$\approx$2.9M) & \textbf{GNN dataset} (n$\approx$72.7K) \\
        \midrule
        Random baseline & 25.2\% & 25.3\% \\
        \textbf{Max free CPU (heuristic)} & \textbf{74.0\%} & \textbf{84.3\%} \\
        Random Forest, Variant D & 65.4\% & --- \\
        GNN prototype & --- & \textbf{65.8\%} \\
        \bottomrule
    \end{tabularx}
\end{table}

\noindent Neither learned model beats the trivial heuristic. The two datasets differ in size and coverage, so heuristic accuracy is not directly comparable across the two columns, but the model-vs-heuristic comparison within each column is valid.

Two findings follow from this table:

\begin{enumerate}
    \item \textbf{Both learned models lose to the heuristic}, on their respective datasets, by a wide margin (8.6 and 18.5 percentage points respectively). Given that reconstructed free CPU is also the dominant feature by importance in the Random Forest (Section~\ref{sec:rf}), this suggests Fuxi's real decisions are substantially explained by a CPU-availability signal that a hand-written rule captures more directly than a model trained to regress toward an arbitrary numeric label.
    \item \textbf{The Random Forest and GNN achieve near-identical Top-1 accuracy (65.4\% vs. 65.8\%) despite the $\sim$40$\times$ data-volume and feature-richness gap.} Read together with Section~\ref{sec:gnn}, this reframes the earlier R² comparison: the GNN's apparently weaker fit is plausibly an artifact of the sample-size and feature asymmetry between the two pipelines, not evidence of an architectural disadvantage --- the two models appear comparably (in)effective once evaluated on the metric that matters.
\end{enumerate}

We attribute the shared shortfall to an \textbf{objective mismatch}: both models are trained with pointwise MSE regression toward the 100/20 label, which does not directly optimize the within-group relative ordering that Top-1 accuracy measures. This is consistent with the learning-to-rank literature's long-standing observation that pointwise regression objectives are a poor proxy for ranking quality~\cite{burges2005, lambda}. A ranking-specific objective (e.g., pairwise margin loss) is the natural next step (Section~\ref{sec:conclusion}).

\subsection{4.5.Overhead and Latency Sensitivity}
\label{sec:latency}

We instrument the plugin with wall-clock timing around the HTTP call to the scoring service, and separately inject artificial delay into the service itself, to characterize production-relevant overhead independent of model quality.

\begin{table}[H]
    \centering
    \caption{Latency sensitivity under injected delay.}
    \begin{tabularx}{0.95\columnwidth}{X X X X X}
        \toprule
        \textbf{Injected delay} & \textbf{Bench mark run time} & \textbf{Pods scheduled} & \textbf{Latency mean} & \textbf{Latency p95} \\
        \midrule
        0ms & 31.2s & 40/40 & 7.8ms & 10.4ms \\
        50ms & 31.2s & 40/40 & 59.0ms & 62.3ms \\
        100ms & 31.6s & 40/40 & 108.9ms & 111.7ms \\
        500ms & 42.7s & 40/40 & 508.8ms & 511.9ms \\
        1500ms & 81.1s & 40/40 & 1508.8ms & 1513.1ms \\
        \bottomrule
    \end{tabularx}
\end{table}

\noindent A fixed $\sim$8--9ms overhead (JSON serialization plus Docker-internal network round-trip) is added on top of the injected delay, independent of its magnitude. All 40 pods scheduled successfully at every tested delay; the plugin's 2-second HTTP timeout was never triggered, leaving only a $\sim$25\% safety margin at the highest tested delay.

\subsection{4.6.Sensitivity to Serving-Container Memory Constraints}
\label{sec:memory}

We separately constrain the scoring service's container memory (via Docker \texttt{mem\_limit}) from 400MB down to 120MB --- below the measured idle baseline of 160.8MB --- while re-running the same 40-pod benchmark on a freshly recreated cluster at each limit.

\begin{table}[H]
    \centering
    \caption{Memory constraint sensitivity.}
    \begin{tabularx}{0.95\columnwidth}{X X X X X}
        \toprule
        \textbf{Memory limit} & \textbf{Pods scheduled} & \textbf{OOM killed} & \textbf{Latency mean} & \textbf{Latency p95} \\
        \midrule
        400MB & 40/40 & No & $\sim$7.9ms & $\sim$10.2ms \\
        250MB & 40/40 & No & $\sim$7.9ms & $\sim$10.2ms \\
        200MB & 40/40 & No & $\sim$7.9ms & $\sim$10.3ms \\
        170MB & 40/40 & No & $\sim$7.8ms & $\sim$11.6ms \\
        \textbf{150MB} & 40/40 & No & $\sim$20.9ms & $\sim$13.3ms \\
        \textbf{120MB} & 40/40 & No & $\sim$33.0ms & $\sim$14.7ms \\
        \bottomrule
    \end{tabularx}
\end{table}

\noindent Averaged across two independent runs; separately confirmed via continuous monitoring that the container operated at $\sim$100\% of the 120MB limit during load, with zero OOM kills.

The service remains fully functional at every tested limit, including 120MB --- below its own idle footprint. Degradation is graceful (rising mean latency, likely garbage-collection pauses under memory pressure) rather than binary (crash/no-crash), with a clear inflection between 170MB and 150MB.

\subsection{4.7.Homogeneous vs. Heterogeneous Workload}
\label{sec:hetero}

Finally, we isolate the effect of workload composition on balancing quality, independent of the scoring model: a homogeneous workload (40 identical \texttt{medium} pods) versus the heterogeneous workload used elsewhere (5 mixed profiles).

\begin{table}[H]
    \centering
    \caption{Homogeneous vs. heterogeneous workload balancing.}
    \begin{tabularx}{0.95\columnwidth}{l X X X}
        \toprule
        \textbf{Workload} & \textbf{Scheduled} & \textbf{CPU (mean /stddev)} & \textbf{Memory (mean /stddev)} \\
        \midrule
        Homogeneous & 40/40 & 41.7\%/ \textbf{1.30\%} &41.7\% / \textbf{1.30\%} \\
        Heterogeneous & 40/40 & 60.8\%/ \textbf{9.01\%} &60.2\% / \textbf{8.93\%} \\
        \bottomrule
    \end{tabularx}
\end{table}

\noindent Balancing stddev under the homogeneous workload is nearly 7$\times$ tighter than under the heterogeneous one.

Notably, this $\sim$7.7-point stddev gap between workload types dwarfs the $\sim$0.3--0.6-point gap observed between baseline and \texttt{AIScore} scheduling in Section~\ref{sec:synthetic} --- a caution against attributing observed balancing differences primarily to the scoring algorithm without controlling for workload composition.

\section{5.Discussion}
\label{sec:discussion} 
\textbf{Regression fit is not a proxy for scheduling quality.} The central empirical result of this paper (Section~\ref{sec:ranking}) is that R², the metric used throughout Sections~\ref{sec:rf}--\ref{sec:gnn} to guide feature engineering, is a poor predictor of the metric that actually governs deployment behavior. A model can show monotonically improving R² across four increasingly sophisticated feature sets (Section~\ref{sec:rf}) while still losing to a one-line heuristic on the ranking task the scheduler actually performs. This mirrors the classical learning-to-rank critique of pointwise objectives~\cite{burges2005, lambda}, now demonstrated concretely in the ML-for-scheduling context. We consider this the paper's primary transferable lesson for ML-for-systems practitioners: when the deployed decision is a ranking/selection among candidates, evaluation should target ranking metrics from the outset, not late in the pipeline.

\textbf{Architecture comparisons can be confounded by resourcing, not capability.} The apparent Random-Forest-over-GNN advantage in Section~\ref{sec:gnn} (R² 0.042 vs. 0.021--0.028) evaporates once both models are evaluated on Top-1 accuracy with awareness of their respective data budgets (Section~\ref{sec:ranking}) --- the GNN achieves comparable ranking quality on $\sim$40$\times$ less data and a far poorer feature set. We do not claim the GNN is \textit{better}, only that the earlier comparison was not apples-to-apples, and caution against drawing architectural conclusions from R² comparisons across pipelines with materially different data volumes.

\textbf{A cheap structural feature rivaled expensive resource-reconstruction features.} \texttt{job\_width} --- free to compute, requiring none of the sweep-line occupancy machinery --- ranked second in feature importance, ahead of a resource feature that took substantial engineering effort to construct correctly (Section~\ref{sec:rf}). This suggests structural/job-level metadata deserves earlier and more thorough exploration in future scheduling-ML work, potentially before investing in fine-grained resource-state reconstruction.

\textbf{Production-facing robustness was better than expected.} Both latency injection (up to 1.5s) and memory constraint (down to 120MB, below idle footprint) produced graceful degradation rather than outright failure (Sections~\ref{sec:latency}--\ref{sec:memory}). This is a reassuring, if secondary, finding for anyone considering an HTTP-served ML scoring plugin in a real deployment, though it was tested only at a single, modest cluster/workload scale.

Relative to Decima~\cite{decima} and DeepRM~\cite{deeprm}, our finding is not that graph-structured or learned representations are unhelpful for scheduling --- Decima's reported gains over heuristics are substantial, and RLScheduler~\cite{rlscheduler} similarly shows learned policies outperforming heuristic priority functions in HPC batch scheduling --- but that isolating the \textit{supervised} sub-problem (would a model reproduce a given historical decision?) from the RL training loop these systems use reveals a metric-selection pitfall that an end-to-end RL evaluation, reporting only downstream job-completion-time improvements, would not surface directly. Decima and DeepRM optimize an RL reward tied directly to job-completion time, sidestepping the pointwise-regression pitfall we identify by construction --- a plausible reason their reported gains over heuristics are more favorable than ours, and indirect support for our proposed fix (Section~\ref{sec:conclusion}) of adopting a ranking- or reward-aligned objective rather than pointwise MSE.

\section{6.Threats to Validity}

\begin{itemize}
    \item Single run per synthetic configuration (Section~\ref{sec:synthetic}) and per parameter value in the sensitivity analyses (Sections~\ref{sec:latency}--\ref{sec:memory}); no confidence intervals from repeated trials.
    \item Domain-restricted negative sampling (Section~\ref{sec:methodology}) is a heuristic approximation of Fuxi's true candidate set at decision time, not verified ground truth.
    \item Memory-related features remain in Alibaba's normalized [0,100] scale; absolute byte values are not recoverable from the trace, limiting direct comparability to the synthetic experiments' byte-based features.
    \item \texttt{pod\_priority} was not extracted from the real trace in any configuration.
    \item The GNN (Sections~\ref{sec:gnn}--\ref{sec:ranking}) used a smaller data sample and a substantially simpler node-feature set than the Random Forest, as discussed in Section~\ref{sec:discussion}; this asymmetry is itself a finding but also limits the strength of any architecture-level claim.
    \item Both models were trained with a pointwise regression objective; the central diagnosis in Section~\ref{sec:ranking} has not yet been validated by retraining with an explicit ranking loss (proposed as future work, Section~\ref{sec:conclusion}).
    \item Sensitivity analyses (Sections~\ref{sec:latency}--\ref{sec:memory}) used a fixed, modest workload (3 nodes, 40 pods); behavior under concurrent, production-scale request volume (many simultaneous \texttt{Score()} calls) was not tested.
    \item All experiments use a single trace (Alibaba Cluster Trace v2018~\cite{alibaba-trace}); generalization to other clusters/traces such as the Google Cluster Trace~\cite{google2011, borg} is untested.
\end{itemize}

\section{7.Conclusion and Future Work}
\label{sec:conclusion}

We built and evaluated \texttt{AIScore}, a learned Kubernetes scheduler score plugin, across a progression from synthetic validation to real-trace training to a rigorous ranking-based evaluation. Our central finding is negative but instructive: despite substantial feature engineering (occupancy reconstruction, failure-domain features, DAG-derived structural features) and a second model architecture (a GraphSAGE-based GNN~\cite{graphsage}), neither learned model surpasses a trivial ``rank by free CPU'' heuristic on the metric that actually reflects scheduling quality --- Top-1 ranking accuracy --- even though both show improving fit under the standard regression metric (R²). We attribute this to a mismatch between the pointwise regression training objective and the ranking-style deployment decision, consistent with the learning-to-rank literature~\cite{burges2005, lambda} and with the RL-based successes of Decima~\cite{decima}, DeepRM~\cite{deeprm}, and RLScheduler~\cite{rlscheduler}, which sidestep this pitfall by optimizing an objective tied to the deployment outcome rather than to an arbitrary pointwise label.

Immediate next steps, several already scoped during this project:

\begin{enumerate}
    \item \textbf{Retrain both models with a ranking-specific loss} (e.g., pairwise margin ranking, following RankNet/LambdaMART~\cite{burges2005, lambda}) to directly test whether closing the objective mismatch closes the gap to the heuristic.
    \item \textbf{Re-run the Random-Forest-vs-GNN comparison at matched data volume and feature richness}, to obtain an architecture comparison unconfounded by the asymmetries noted in Section~\ref{sec:discussion}.
    \item \textbf{Investigate the \texttt{job\_width} finding directly} (e.g., stratifying by job-width buckets) to understand the underlying placement pattern it captures, and to inform reward design for an eventual RL-based scheduler.
    \item \textbf{Extend the deep experimental evaluation} with an oracle baseline (post-hoc optimal placement via a combinatorial solver), a makespan metric, dynamic/wave-arrival workloads replaying real trace timestamps, and node-failure injection --- components scoped but not yet executed in this project.
    \item \textbf{Repeat key experiments with confidence intervals} across multiple random seeds, and validate findings against a second cluster trace such as the Google Cluster Trace~\cite{google2011, borg}.
\end{enumerate}

All code --- the plugin, training pipelines, evaluation scripts, and sensitivity-analysis harnesses --- is available for reproduction in the accompanying artifact repository.
\section*{Appendix: Reproducibility}

Data source: Alibaba Cluster Trace v2018, \texttt{alibaba/clusterdata} repository (\texttt{cluster-trace-v2018})~\cite{alibaba-trace}.

\begin{lstlisting}[language=bash, breaklines=true]
	git clone https://github.com/1901asyl/aiscore-scheduler
	cd aiscore-scheduler
	
	# Synthetic sanity check (Section 4.1)
	python3 benchmark/run_benchmark.py --label baseline2 --n 8
	python3 benchmark/run_benchmark.py --label aiscore2 --n 8
	python3 benchmark/run_benchmark.py --compare baseline2 aiscore2
	
	# Random Forest on real trace (Section 4.2)
	python3 data/build_training_set.py
	python3 ml-server/train.py
	
	# GNN prototype (Section 4.3)
	python3 data/build_job_graphs.py
	python3 data/build_training_set_gnn.py
	python3 ml-server/train_gnn.py
	
	# Ranking evaluation (Section 4.4)
	python3 data/evaluate_ranking.py
	python3 data/evaluate_ranking_gnn.py
	
	# Latency sensitivity (Section 4.5)
	bash run_latency_experiment.sh
	
	# Memory sensitivity (Section 4.6)
	bash run_memory_experiment.sh
	
	# Homogeneous vs. heterogeneous workload (Section 4.7)
	python3 benchmark/run_benchmark.py --label homogeneous_medium --n 8 --homogeneous medium
	python3 benchmark/run_benchmark.py --label heterogeneous_mixed --n 8
	python3 benchmark/run_benchmark.py --compare homogeneous_medium heterogeneous_mixed
\end{lstlisting}

All code --- the plugin, training pipelines, evaluation scripts, and sensitivity-analysis harnesses --- is publicly available at \url{https://github.com/1901asyl/aiscore-scheduler} (commit ab53a9f).


\end{document}